\documentclass[a4paper,11pt]{article}
\usepackage{pos}
\usepackage{amsmath}
\usepackage{graphicx}
\usepackage{subcaption}

\title{Diagonal Kenney-Laub Rational Approximation to the Overlap Dirac Operator}
\ShortTitle{Diagonal KL Rational Approximation}

\author[a, b]{Stephan D\"urr}
\author*[c]{Stylianos Gregoriou}
\author[c]{Giannis Koutsou}

\affiliation[a]{University of Wuppertal,\\
Gau{\ss}stra{\ss}e 20, 42119 Wuppertal, Germany}

\affiliation[b]{JSC, Forschungszentrum J\"ulich,\\
52425 J\"ulich, Germany}

\affiliation[c]{The Cyprus Institute,\\
20 Kavafi Street, Nicosia 2121, Cyprus}

\emailAdd{s.durr$\,($AT$)\,$fz$-$juelich.de}
\emailAdd{s.gregoriou$\,($AT$)\,$cyi.ac.cy}
\emailAdd{g.koutsou$\,($AT$)\,$cyi.ac.cy}

\abstract{We propose a practical formulation of the overlap Dirac operator in lattice QCD that employs the diagonal Kenney-Laub rational iterates -- expressed via their partial fraction decomposition -- to approximate the matrix sign function. We investigate this approximation using the Brillouin operator as kernel, in addition to the standard Wilson Dirac operator. Numerical results show improved chiral symmetry preservation and computational efficiency compared to the Chebyshev polynomial approach.}

\FullConference{The 42nd International Symposium on Lattice Field Theory (LATTICE2025)\\
2--8 November 2025\\
Tata Institute of Fundamental Research, Mumbai, India}

\begin{document}

    \maketitle
    
    \section{Introduction}
        
        The Wilson term in the Wilson-Dirac fermion discretization successfully removes fermion doublers but explicitly breaks the chiral symmetry of continuum QCD, consistent with the Nielsen-Ninomiya no-go theorem. Ginsparg and Wilson~\cite{ginsparg1982remnant-48c} found a way to reconcile these requirements by redefining chiral symmetry on the lattice. A Dirac operator $D$ is chiral within this definition if it satisfies the Ginsparg-Wilson (GW) relation:
        \begin{align}
            \{ \gamma_5, D \} = a D \gamma_5 D \, .
            \label{eq:GW}
        \end{align}
        where $a$ is the lattice spacing. Two types of undoubled fermion formulations satisfy this relation exactly, namely the domain wall fermions in the limit where the extent of the fifth dimension goes to infinity ($N_5\to\infty$) and overlap fermions. In this work, we focus on overlap fermions.
        
        The massive overlap operator, proposed by Neuberger and Narayanan~\cite{neuberger1998exactly-7eb,neuberger1998more-aea}, takes the form:
        \begin{align}
            aD^{\mathrm{ov}}_m = \left(\rho + \frac{am}{2}\right)\mathbb{I} + \left(\rho - \frac{am}{2}\right)\gamma_5 \, \mathrm{sgn}[\mathbb{X}] \, ,
            \label{eq:overlap}
        \end{align}
        with $\mathbb{X} = \gamma_5(aD^{\mathrm{ker}}_0 - \rho\mathbb{I})$, where $D^{\mathrm{ker}}_0$ is a massless Dirac operator serving as the kernel and $\rho$ is a dimensionless parameter with $0 < \rho < 2$. The matrix sign function for Hermitian matrix $\mathbb{H}$ can be expressed as $\mathrm{sgn}[\mathbb{H}] = \mathbb{H}(\mathbb{H}^2)^{-1/2}$, and approximating this with sufficient precision is the main computational challenge for overlap fermions.
        
        Our approach explores two complementary directions: the \emph{diagonal Kenney-Laub iterates} as a rational approximation to the sign function, and the \emph{Brillouin operator} as a chirally-improved kernel. In the following sections, we examine the theoretical properties of each component and present numerical results assessing the practical advantages of this combination.
        
    \section{The Diagonal Kenney-Laub Method}
        
        In the early 1990s, Kenney and Laub~\cite{10.1137/0612020} published a rational recursive scheme for approximating the matrix sign function. Starting from an indefinite Hermitian matrix $\mathbb{H}$ with no zero eigenvalues, a sequence is built recursively that converges to $\mathrm{sgn}[\mathbb{H}]$:
        \begin{align}
            \mathrm{sgn}[\mathbb{H}] = \lim_{k \to \infty} \mathbb{H}_k \, , \quad \text{with} \quad \mathbb{H}_{k+1} = f_{mn}(\mathbb{H}_k) \quad \text{and} \quad \mathbb{H}_0 = \mathbb{H} \, .
        \end{align}
        The $(m,n)$ Kenney-Laub (KL) iterates $f_{mn}$ are constructed from Pad\'e approximants of $(1-t)^{-1/2}$, $t \equiv 1 - x^2$, where $m$ and $n$ are the degrees of the numerator and denominator polynomials. We focus on the \emph{diagonal} case ($m = n$), using a single recursion ($k=1$) with varying order $n$~\cite{10.48550/arxiv.1701.00726}:
        \begin{align}
            \mathrm{sgn}[\mathbb{H}] \approx f_{nn}(\mathbb{H}) = \mathbb{H} \frac{\sum_{i=0}^{n} \binom{2n+1}{2i+1} (\mathbb{H}^2)^i}{\sum_{i=0}^{n} \binom{2n+1}{2i} (\mathbb{H}^2)^i} \, .
            \label{eq:KL_diagonal}
        \end{align}
        
\begin{figure}[t]
            \centering
            \begin{subfigure}{0.5\textwidth}
                \centering
                \includegraphics[width=\linewidth]{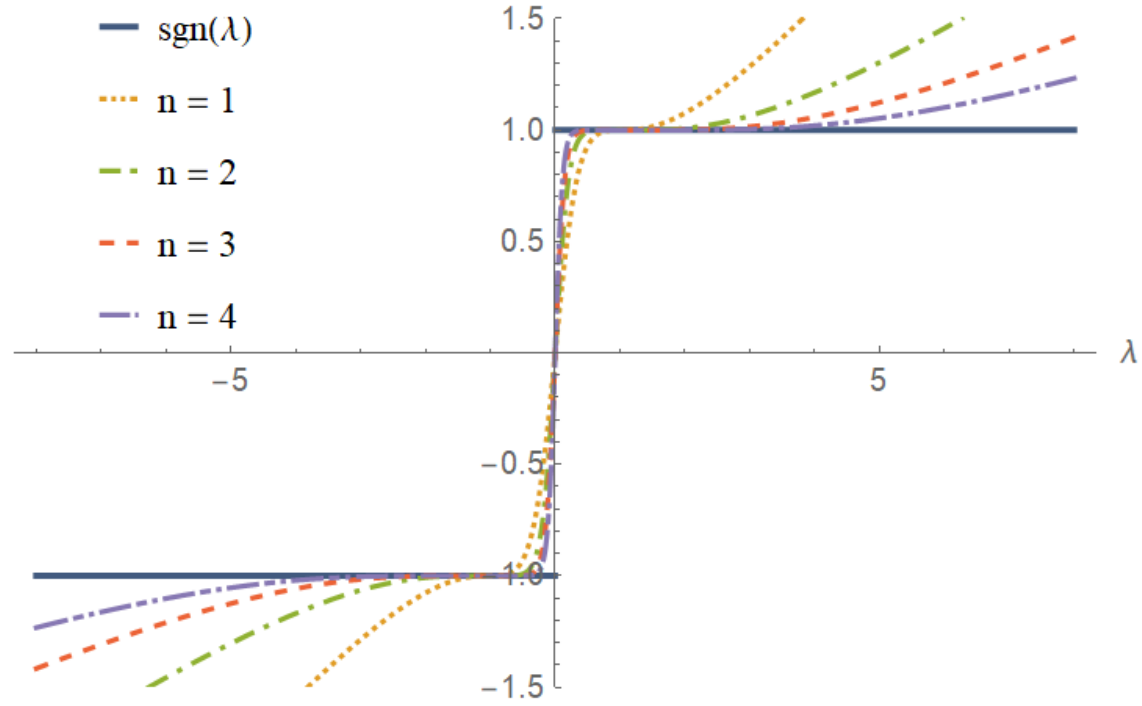}
                \caption{Diagonal KL method: $n = 1, 2, 3, 4$.}
            \end{subfigure}
            \begin{subfigure}{0.48\textwidth}
                \centering
                \includegraphics[width=\linewidth]{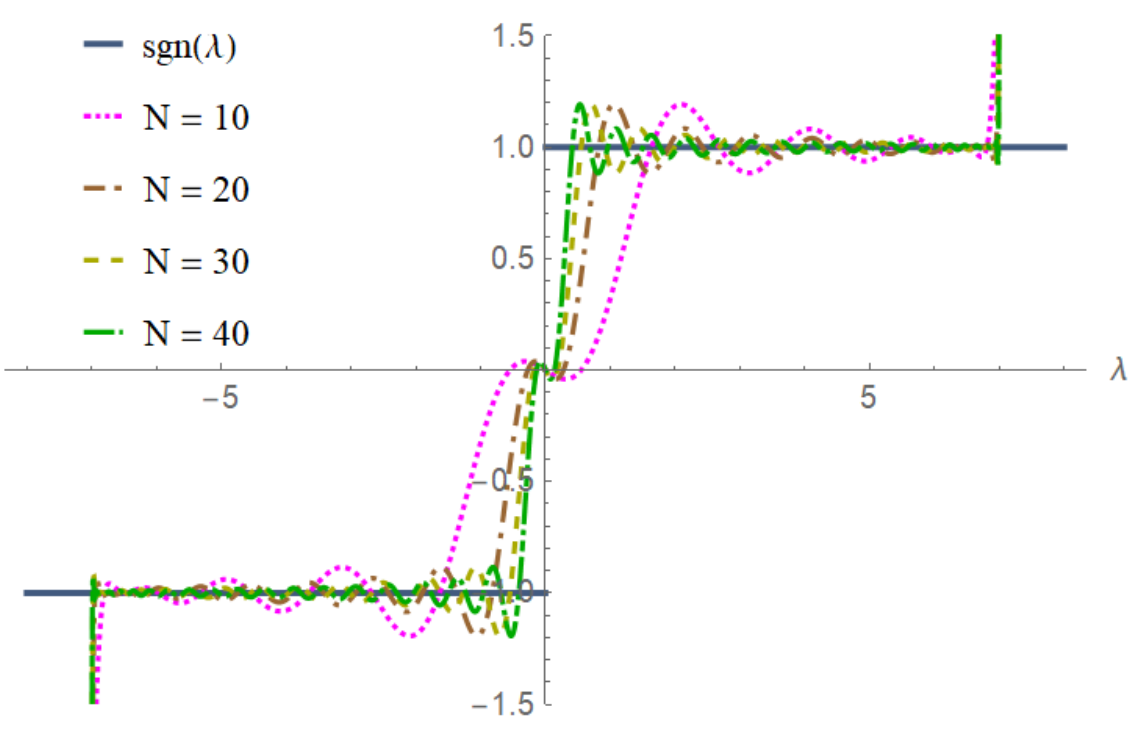}
                \caption{Chebyshev method: $N = 10, 20, 30, 40$.}
            \end{subfigure}
            \caption{Scalar sign function (solid blue line) compared to approximations using diagonal KL iterates, defined on all of $\mathbb{R}$ (left), and Chebyshev polynomials, defined on $|\lambda| \in [0.01, 7.0]$ (right). The latter interval approximately spans the eigenvalue range for a Wilson kernel in 4D with $\rho=1$.}
            \label{fig:scalar_comparison}
        \end{figure}

        Fig.~\ref{fig:scalar_comparison} illustrates how the diagonal KL iterates approximate the scalar sign function, compared to Chebyshev polynomials. The KL method exhibits smooth, monotonic convergence that improves uniformly with increasing $n$ across all of $\mathbb{R}$, while Chebyshev, where $2N$ is approximately the polynomial degree, shows oscillatory behavior with less consistent pointwise improvement -- particularly near zero -- and diverges outside its prescribed, bounded domain of approximation.
        
        The diagonal KL iterates possess several key advantages. Most importantly, they exhibit \emph{global convergence}~\cite{10.1137/0612020}: no eigenvalue estimation or matrix rescaling of $\mathbb{X}$ is required, unlike other approximation schemes. This eliminates a computational step that is necessary for methods such as Chebyshev or Zolotarev, and which needs to be carried out every time the background gauge field configuration changes.
        
        Furthermore, the diagonal KL iterates admit a numerically convenient
        partial fraction decomposition:
        \begin{align}
            f_{nn}(\mathbb{H}) = \mathbb{H} \left( c_0 + \sum_{i=1}^{n} \frac{c_i}{\mathbb{H}^2 + \sigma_i} \right) \, ,
            \label{eq:partial_fraction}
        \end{align}
        where all coefficients $c_i > 0$ and shifts $\sigma_i > 0$ are trigonometric functions that depend \emph{only} on $n$:
        \begin{align}
            c_0 = \frac{1}{2n+1}\, , \quad c_i = \frac{2c_0}{\cos^2\left[\frac{(2i-1)\pi}{4n+2}\right]}\, , \quad \sigma_i = \tan^2\left[\frac{(2i-1)\pi}{4n+2}\right]\, .
        \end{align}
        This form is particularly well-suited for Multi-Shift Conjugate Gradient (MSCG) solvers, which can solve all shifted systems $(\mathbb{H}^2 + \sigma_i)x = b$ for a set
        $0<\sigma_1<\sigma_2<\ldots<\sigma_n$ simultaneously, at approximately the cost of the smallest, and thus slowest-converging, shift $\sigma_1$.
        
    \section{The Brillouin Operator}
        
        The Brillouin operator, developed in Ref.~\cite{10.1103/physrevd.83.114512}, is a parameter-free Wilson-type operator that maintains the Wilson structure but employs extended stencils for its covariant derivative and gauged Laplacian. Specifically, it uses a 64-point stencil for the isotropic covariant derivative and an 81-point stencil for the Brillouin gauged Laplacian, compared to the 2-point and 9-point stencils of standard Wilson action. While this increases the application cost by a factor of 3-4 per lattice site, the hops stay within the 4D hypercube, keeping the overhead controllable.
        
\begin{figure}[t]
            \centering
            \begin{subfigure}{0.49\textwidth}
                \centering
                \includegraphics[width=\linewidth]{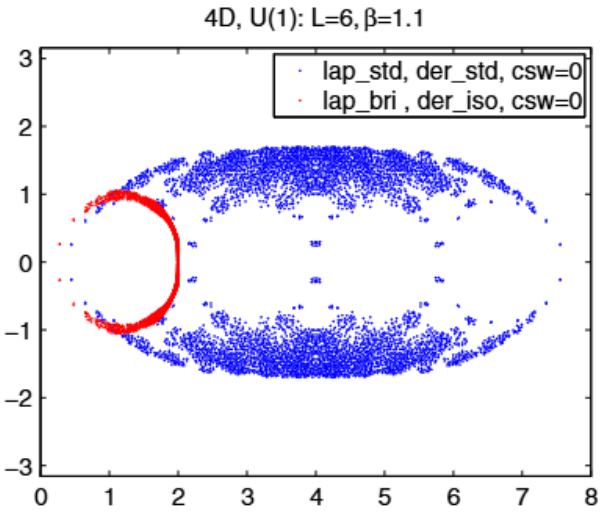}
            \end{subfigure}
            \begin{subfigure}{0.49\textwidth}
                \centering
                \includegraphics[width=\linewidth]{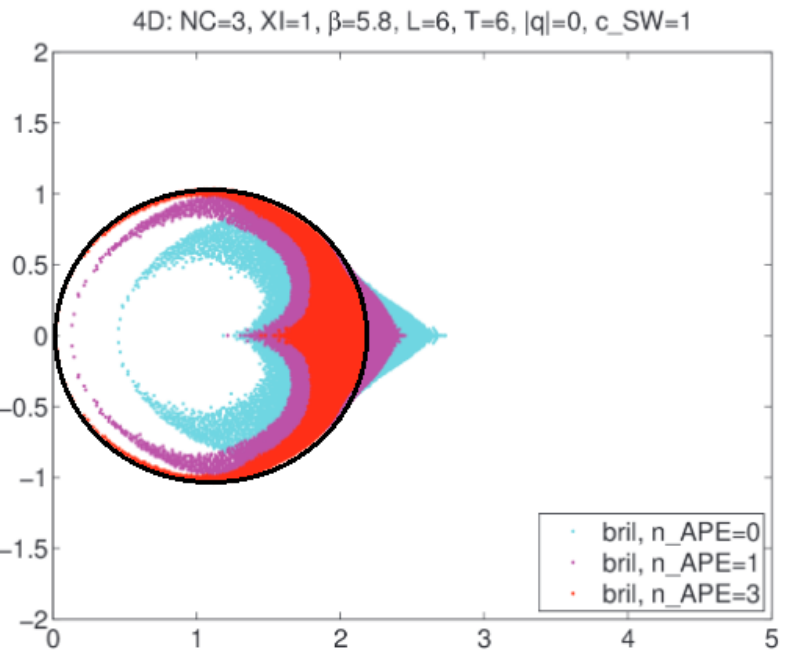}
            \end{subfigure}
            \caption{Eigenvalue spectra for the Wilson (left) and Brillouin (right) operators on thermalized configurations. The Brillouin eigenvalue spectrum is compared to the ideal GW form (black circle) for increasing number of APE smearing steps as indicated in the legend. Figures adapted from Ref.~\cite{10.1103/physrevd.83.114512} (left) and Ref.~\cite{10.48550/arxiv.1701.00726} (right).}
            \label{fig:spectral}
        \end{figure}

        The design goal of the Brillouin operator is to reduce the degree of rotational symmetry breaking near the boundaries of the Brillouin zone, hence its name. For our purposes, its most important advantage is in its spectral properties. As shown in Fig.~\ref{fig:spectral}, the Brillouin operator exhibits smaller condition numbers than the Wilson operator. More importantly, its eigenvalue spectrum is much closer to the ideal shifted-circular spectrum that one expects from the Ginsparg-Wilson relation, making it a promising candidate as an overlap kernel.
        
    \section{Numerical Results}

        \subsection{Setup}
            
            We performed proof-of-concept calculations on $48 \times 24^3$ lattices at $\beta_{\mathrm{QCD}} = 6.20$ ($a \approx 0.066$~fm) using quenched gauge field configurations with one APE smearing step with $\alpha_{\mathrm{APE}} = 0.72$. For this proof-of-concept, we refrain from tuning any parameters, keeping the setup minimal with $\rho = 1$ and $c_{SW} = 0$. We compare two kernels, namely Brillouin vs.\ Wilson, and two approximation methods, the diagonal KL rational vs.\ Chebyshev polynomial, evaluating performance across several metrics.
            
        \subsection{Ginsparg-Wilson Violation}
            
\begin{figure}[t]
                \centering
                \begin{subfigure}{0.49\textwidth}
                    \centering
                    \includegraphics[width=\linewidth]{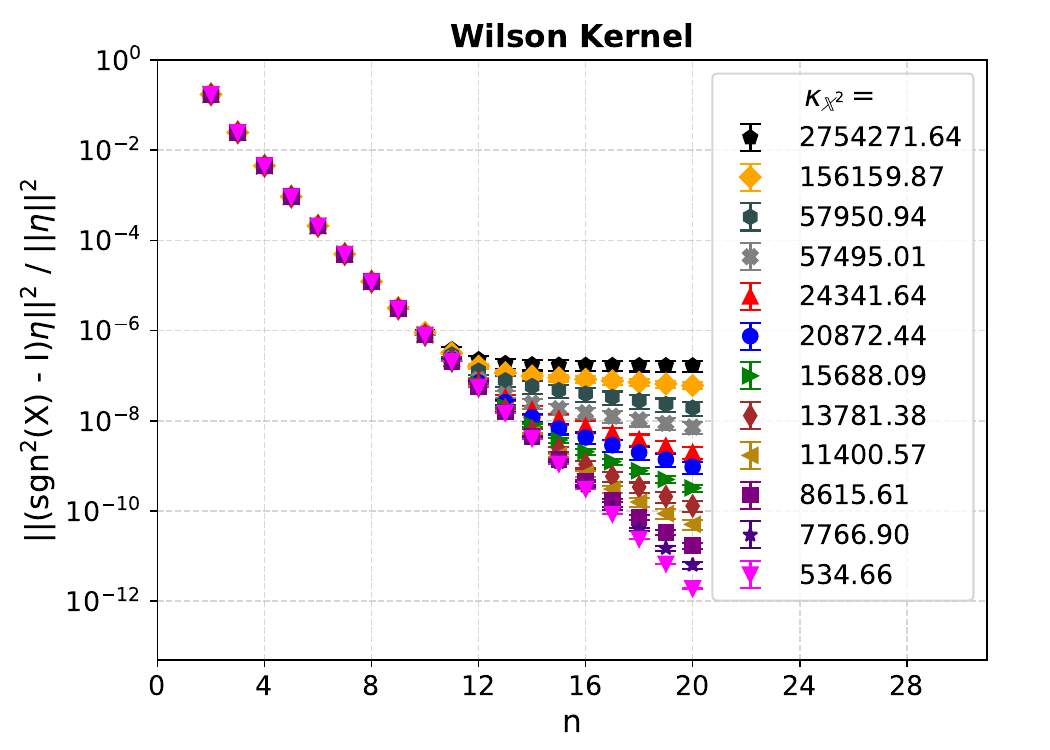}
                \end{subfigure}
                \begin{subfigure}{0.49\textwidth}
                    \centering
                    \includegraphics[width=\linewidth]{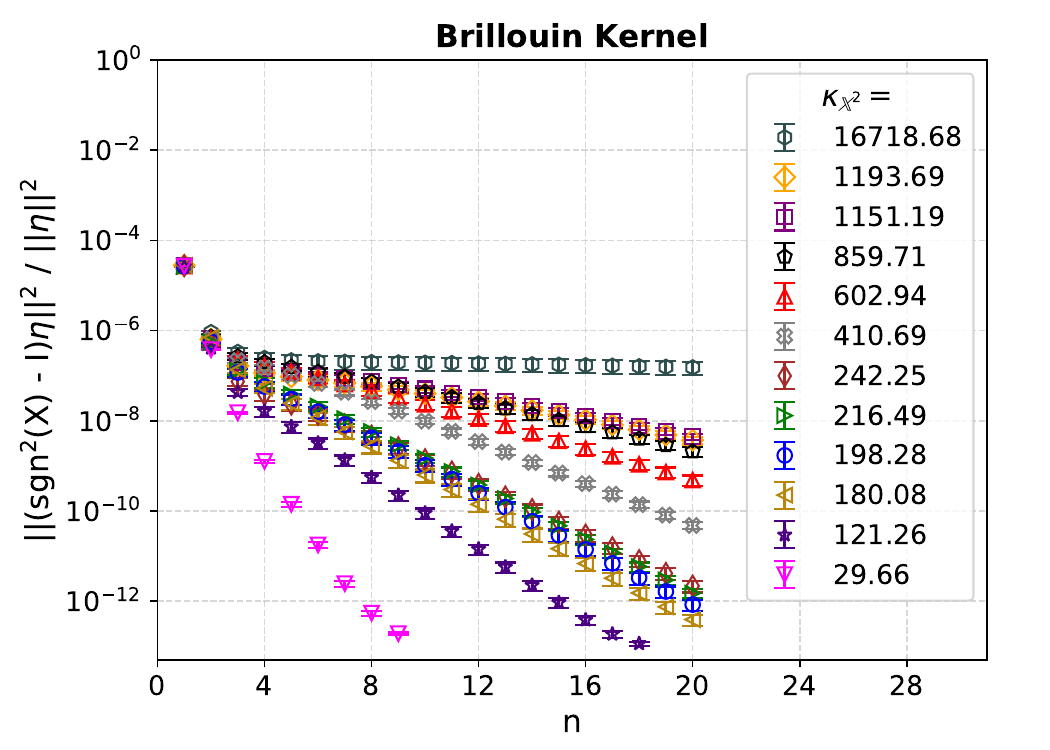}
                \end{subfigure}
                \caption{GW violation vs. diagonal KL order for the Wilson (left) and Brillouin (right) kernels. Same color and symbol combination refers to the same configuration, whose condition number $\kappa_{\mathbb{X}^2}$ for that kernel is given in the respective legend.}
                \label{fig:GW_violation}
            \end{figure}

            For the (dimensionless) massless overlap operator $aD^{\mathrm{ov}} = \rho(\mathbb{I} +
            \gamma_5\,\mathrm{sgn}[\mathbb{X}])$, we define the Ginsparg-Wilson violation ($\Delta_{\rm GW}$) with $\rho = 1$:
            \begin{align}
                \Delta_{\mathrm{GW}} = \gamma_5 aD^{\mathrm{ov}} + aD^{\mathrm{ov}} \gamma_5 - aD^{\mathrm{ov}} \gamma_5 aD^{\mathrm{ov}} \, ,
            \end{align}
            which vanishes for an exact sign function. Fig.~\ref{fig:GW_violation} shows the averaged norm $||\Delta_{\mathrm{GW}}\eta||^2/||\eta||^2$ over random vectors $\eta$, as a function of diagonal KL order $n$ for several gauge configurations with different condition numbers $\kappa_{\mathbb{X}^2}$.
            
            We observe systematic, monotonic improvement with increasing KL order $n$,  as expected from theory. The Brillouin kernel consistently outperforms the Wilson kernel, requiring a lower KL order $n$ to achieve the same $||\Delta_{\mathrm{GW}}\eta||^2/||\eta||^2$. Note also that for a given gauge configuration (same shape and color between left and right panels), the Brillouin kernel exhibits a smaller condition number, as expected.
            
        \subsection{Physical Observables}

\begin{figure}[t]
                \centering
                \begin{subfigure}{0.49\textwidth}
                    \centering
                    \includegraphics[width=\linewidth]{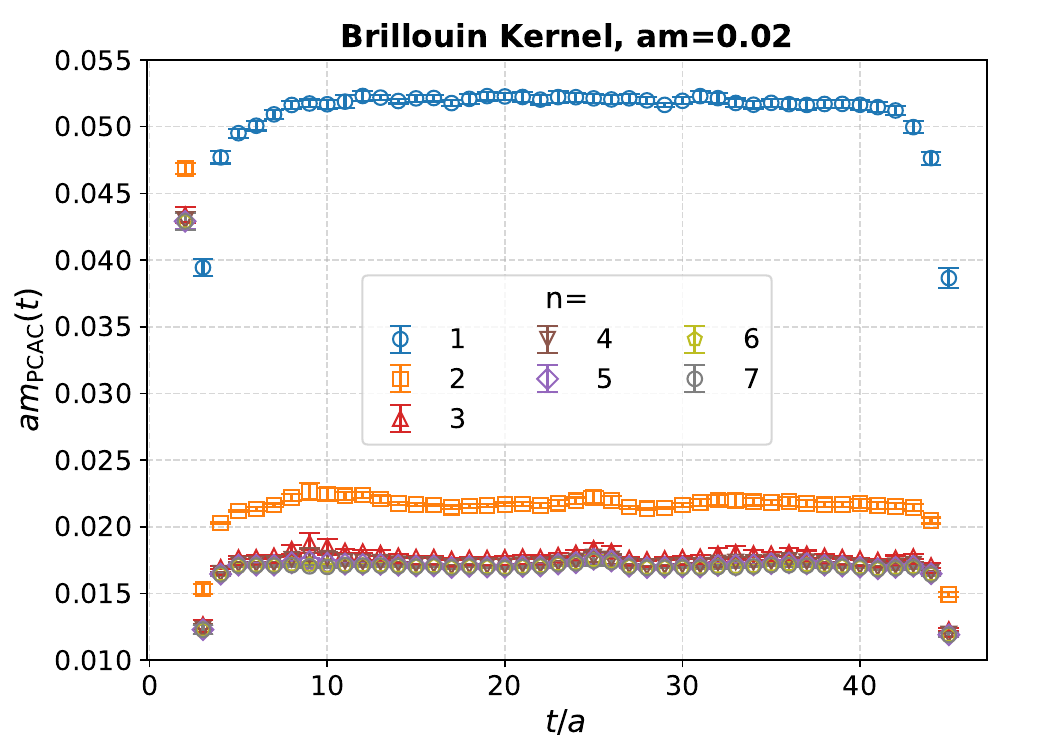}
                \end{subfigure}
                \begin{subfigure}{0.49\textwidth}
                    \centering
                    \includegraphics[width=\linewidth]{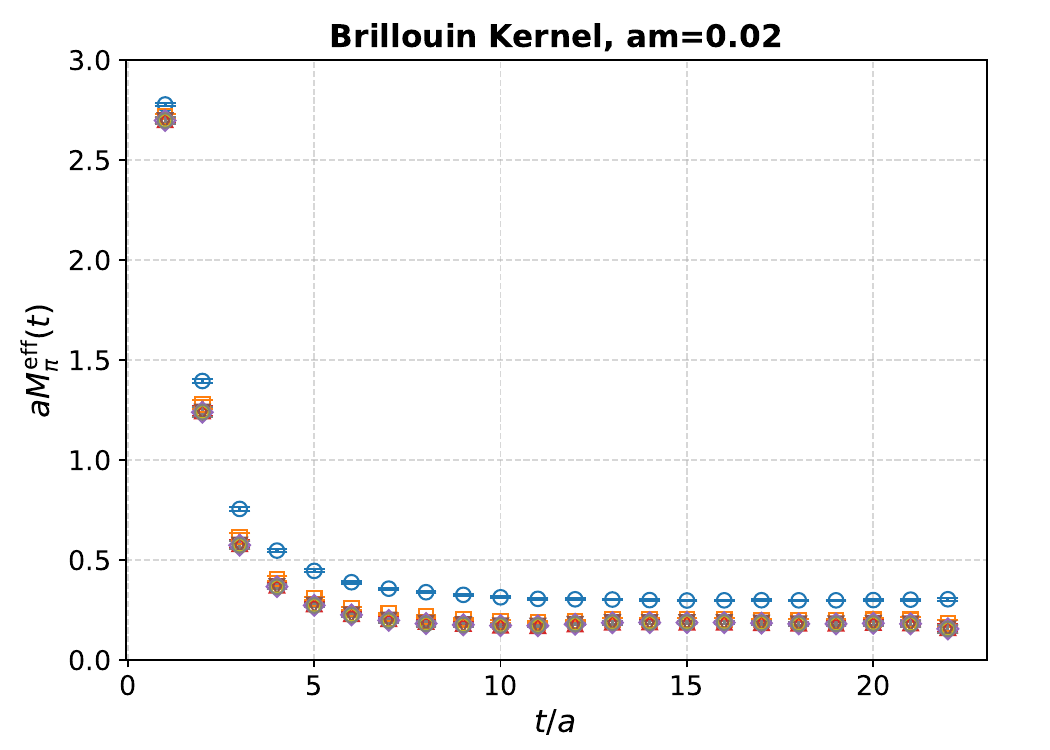}
                \end{subfigure}
                \caption{PCAC mass (left) and pion effective mass (right) for the Brillouin kernel at $\rho=1$ and $am = 0.02$ with diagonal KL orders $n = 1, \ldots, 7$. The number of configurations is $O(20)$.}
                \label{fig:PCAC_pion}
            \end{figure}

            We evaluate the PCAC mass, as defined in Ref.~\cite{Durr:2011ed}, as well as pion two-point correlation functions varying the bare quark mass ($am$) of the overlap operator (see Eq.~\ref{eq:overlap}). Fig.~\ref{fig:PCAC_pion} shows a representative example of the PCAC mass $am_{\mathrm{PCAC}}(t)$ and the pion effective mass $aM_\pi^\mathrm{eff}(t)$ as a function of the Euclidean time separation for the Brillouin kernel at $am = 0.02$ for various KL orders $n$. As can be seen, excited-state effects are sufficiently suppressed starting at around $t/a=12$, where we identify a plateau region. Furthermore, we observe a clear convergence in both the PCAC mass and pion mass with increasing $n$.
                
\begin{figure}[t]
                \centering
                \begin{subfigure}{0.49\textwidth}
                    \centering
                    \includegraphics[width=\linewidth]{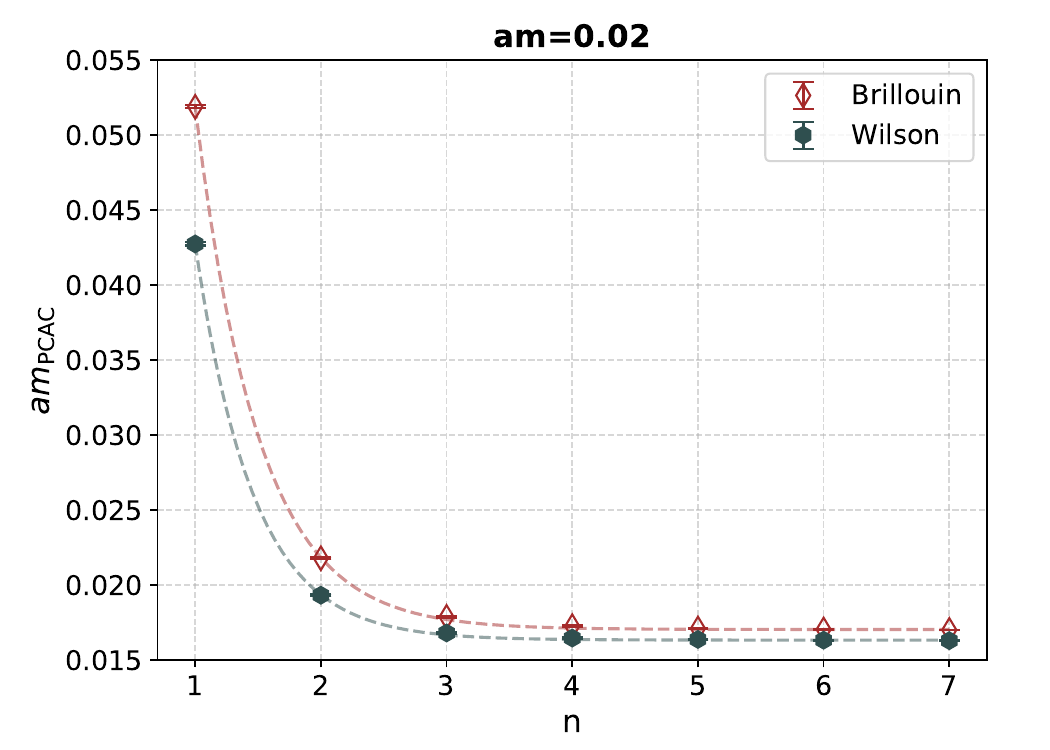}
                \end{subfigure}
                \begin{subfigure}{0.49\textwidth}
                    \centering
                    \includegraphics[width=\linewidth]{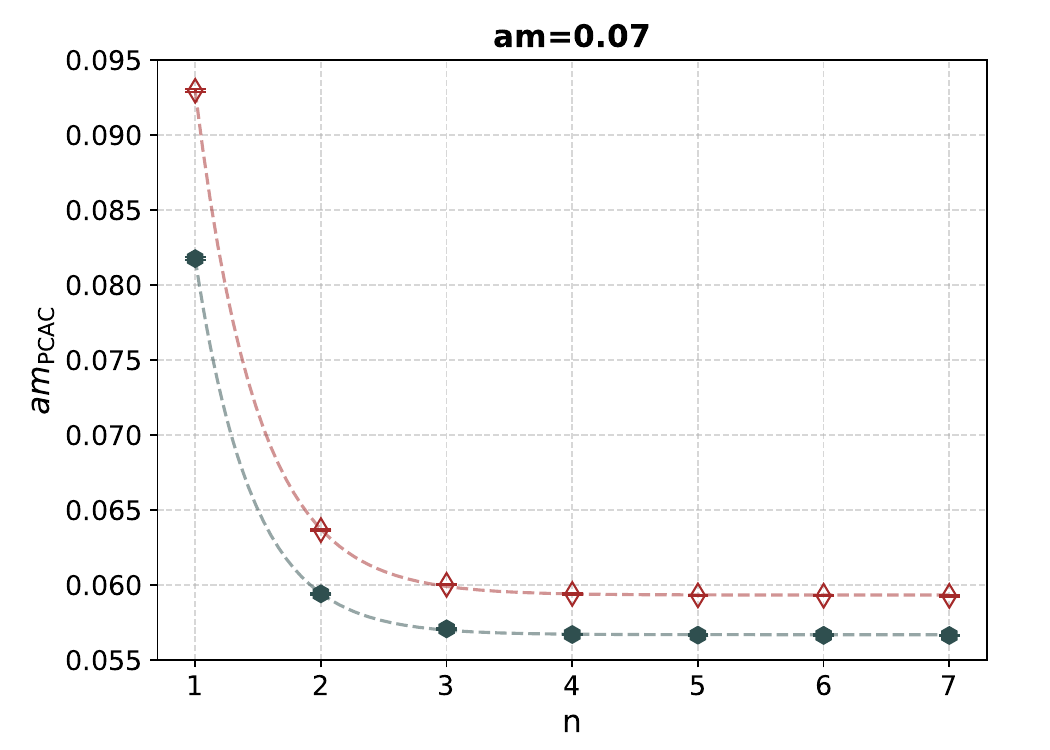}
                \end{subfigure}
                \caption{Plateau PCAC mass vs.\ diagonal KL order for $am = 0.02$ (left) and $am = 0.07$ (right), comparing Wilson (filled) and Brillouin (empty) kernels. The statistics  are the same as in Fig.~\ref{fig:PCAC_pion}.}
                \label{fig:PCAC_convergence}
            \end{figure}

            We fit the PCAC mass in the identified plateau region, obtaining the values shown in Fig.~\ref{fig:PCAC_convergence}, plotted against the KL order for two different bare mass values. Both kernels exhibit rapid exponential convergence, reaching their asymptotic values within just a few approximation orders. Beyond $n\simeq5$ we see no improvement within the statistics at hand. At convergence, the two kernels yield different PCAC mass values for a given $am$, consistent with expectations.
            
        \subsection{Critical Bare Mass}
                
\begin{figure}[t]
                \centering
                \begin{subfigure}{0.49\textwidth}
                    \centering
                    \includegraphics[width=\linewidth]{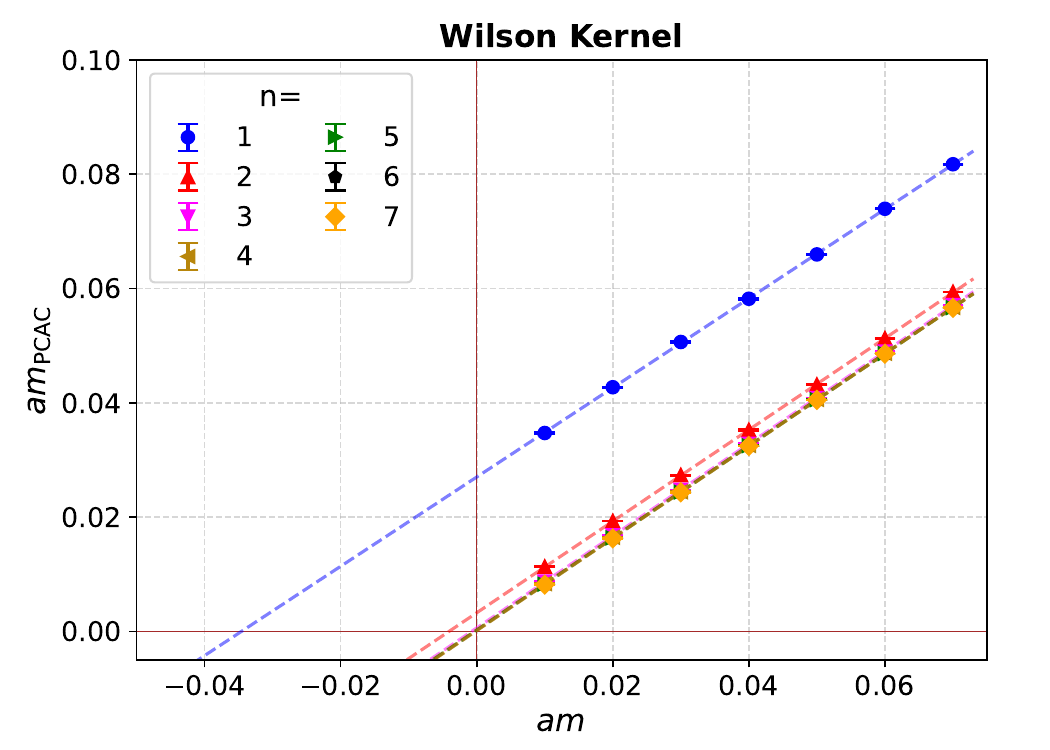}
                \end{subfigure}
                \begin{subfigure}{0.49\textwidth}
                    \centering
                    \includegraphics[width=\linewidth]{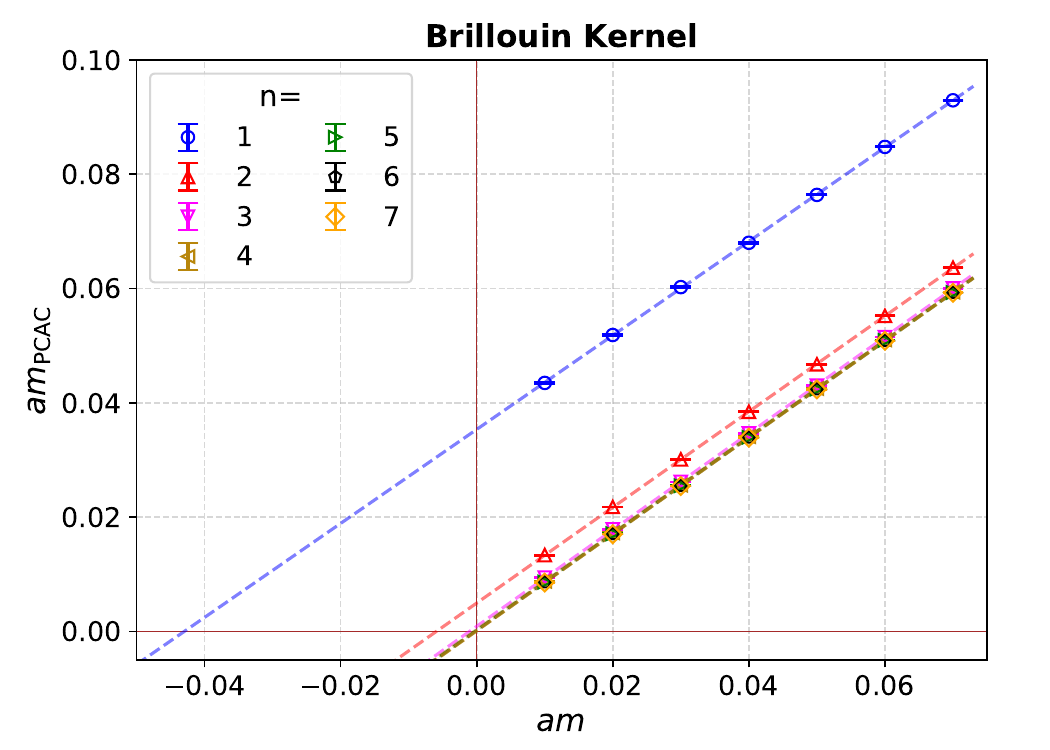}
                \end{subfigure}
                \caption{Plateau PCAC mass vs.\ bare mass for Wilson (left) and Brillouin (right) kernels at different diagonal KL orders. The dashed lines result from linear fits to the data shown. The critical bare mass is the x-intercept.  The vertical line shows $am=0$ for reference. The statistics are the same as in Fig.~\ref{fig:PCAC_pion}.}
                \label{fig:critical_mass}
            \end{figure}

            In Fig.~\ref{fig:critical_mass} we plot the fitted PCAC mass as a function of the bare mass. The critical bare mass $am_{\mathrm{bare}}^{\mathrm{critical}}$, defined as the value of the bare mass for which $am_{\rm PCAC}=0$, is obtained via linear fits to this data.  $am_{\mathrm{bare}}^{\mathrm{critical}}$ is a direct measure of the quality of the chiral symmetry achieved; a perfectly chiral operator has zero critical bare mass. As can be seen in Fig.~\ref{fig:critical_mass}, both kernels show systematic improvement with increasing order $n$. The critical bare mass $am_{\mathrm{bare}}^{\mathrm{critical}}$ tends to zero for $n\to\infty$, but it is consistent with zero (within errors) for finite $n$.
                
        \subsection{Comparison with Chebyshev polynomial method}
                
\begin{figure}[t]
                \centering
                \begin{subfigure}{0.49\textwidth}
                    \centering
                    \includegraphics[width=\linewidth]{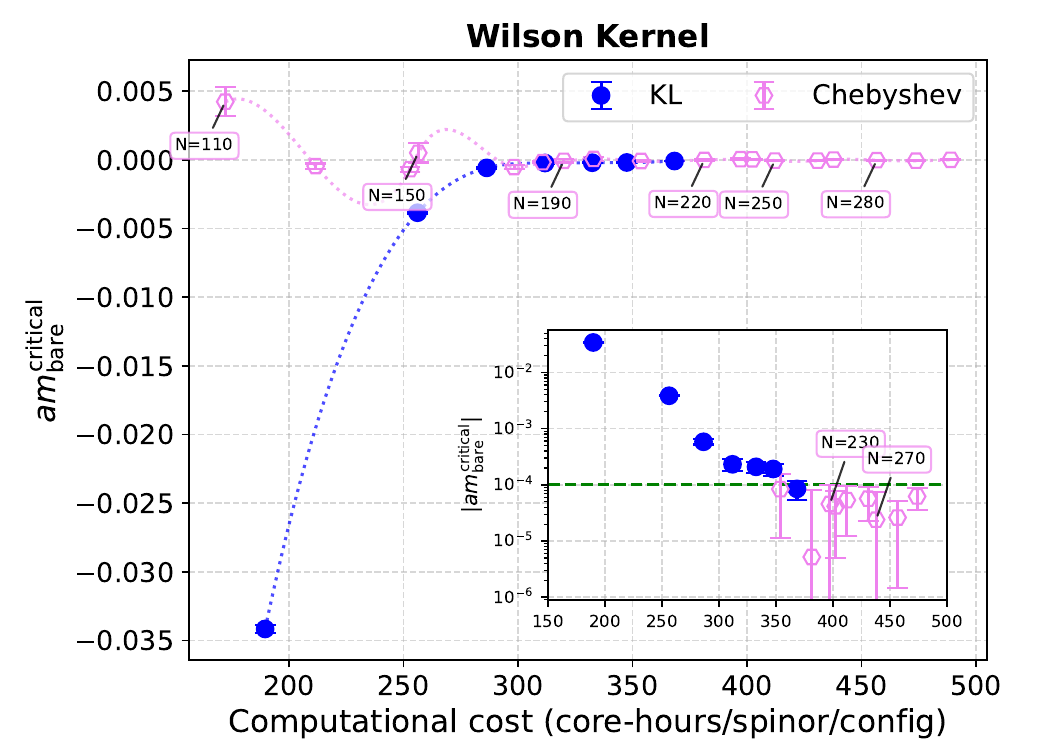}
                \end{subfigure}
                \begin{subfigure}{0.49\textwidth}
                    \centering
                    \includegraphics[width=\linewidth]{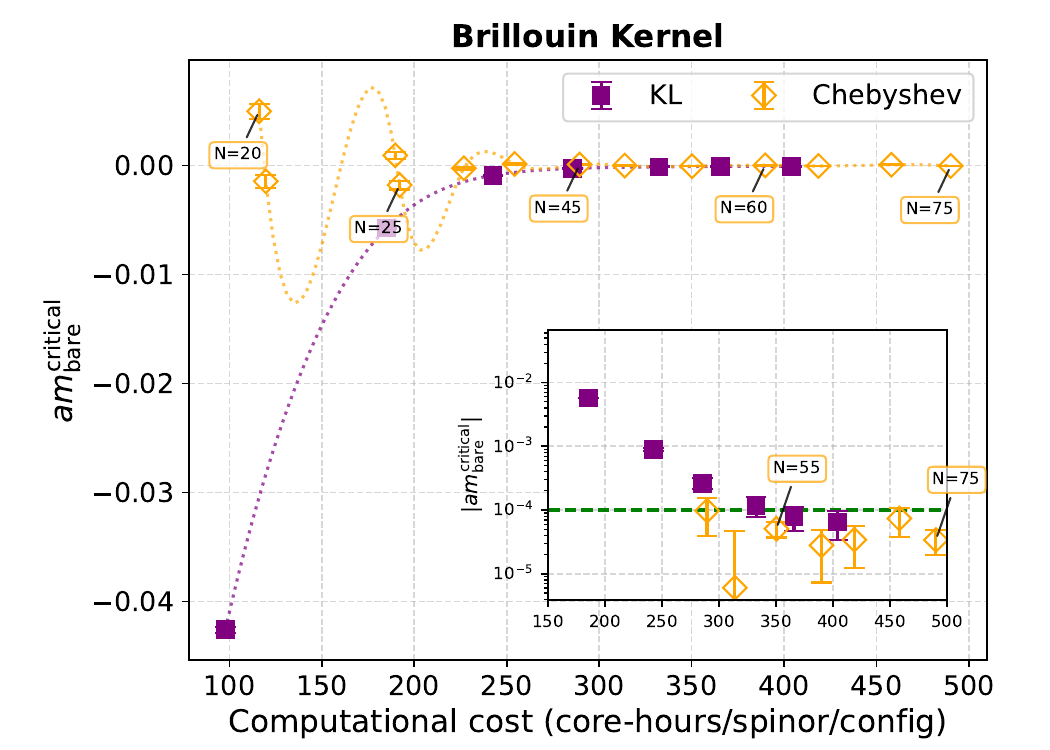}
                \end{subfigure}
                \caption{Critical bare mass obtained
                  vs.\ computational cost as defined in the text, for Wilson (left) and Brillouin (right) kernels, comparing the KL (filled) and Chebyshev (empty) methods.  For KL, we show data for $n=1$ to $n=7$ from left to right, while for Chebyshev we select representative values of the polynomial order $N$ with values indicated in the plot. Insets show a subset of the data with semi-log scale; the green line marks the target precision of $|am_{\mathrm{bare}}^{\mathrm{critical}}| = 10^{-4}$.}
                \label{fig:cost_comparison}
            \end{figure}

            To compare the computational requirements of our overlap procedure among different approximation orders, we estimate, for each case, the core-hours needed to obtain one spinor on one configuration for a bare overlap quark mass, $am$, such that it yields a pion mass of around 300~MeV. In Fig.~\ref{fig:cost_comparison} we plot $am_{\rm bare}^{\rm critical}$ against this cost for different KL and Chebyshev orders. The KL method shows predictable, monotonic convergence, while the Chebyshev method oscillates before stabilizing. This makes KL more controllable when targeting a specific precision.
                
\begin{figure}[t]
                \centering
                \includegraphics[width=0.55\textwidth]{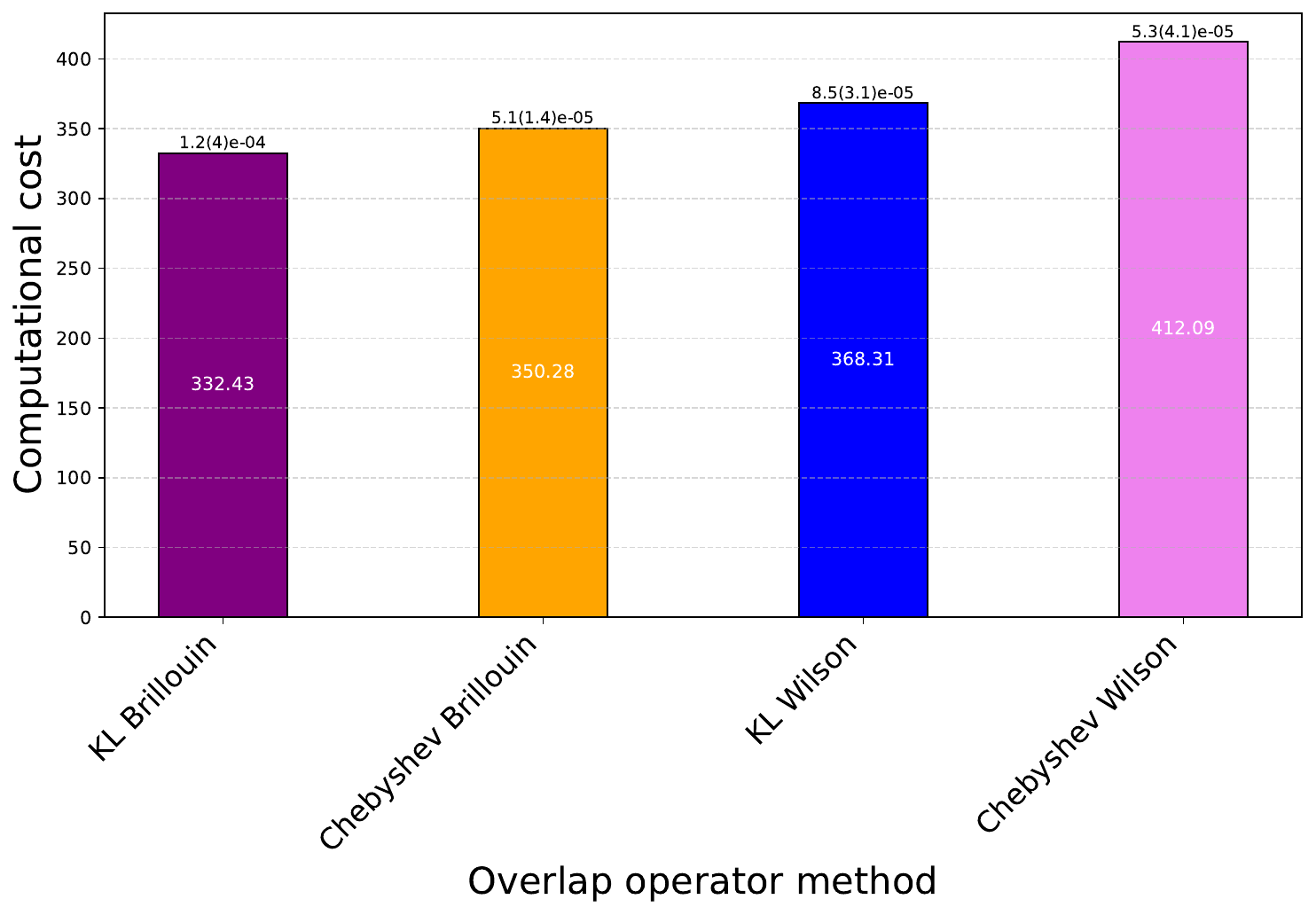}
                \caption{Computational cost for the four method-kernel combinations to reach $|am_{\mathrm{bare}}^{\mathrm{critical}}| \simeq 10^{-4}$. The achieved precision is indicated above each bar.}
                \label{fig:summary}
            \end{figure}

            In Fig.~\ref{fig:summary} we summarize the computational
            cost required to reach the target precision
            $|am_{\mathrm{bare}}^{\mathrm{critical}}| < 10^{-4}$. The
            KL-Brillouin combination requires computational cost
            comparable to - or slightly less than - the
            Chebyshev-Wilson combination, while offering more
            predictable convergence.
            
    \section{Conclusions}
        
        We have presented an alternative approach to implementing overlap fermions by combining the diagonal Kenney-Laub rational approximation with the Brillouin operator as the kernel. Key features of this combination include the elimination of the requirement to estimate the eigenvalues of the kernel operator, improved kernel conditioning, predictable monotonic convergence with increasing approximation order, and minimal parameter tuning.
        
        Our proof-of-concept tests find an overall performance that is comparable to the established Chebyshev-Wilson combination, with indications of potential cost advantages for reaching a specific precision in the overlap approximation. Further details will be presented in a forthcoming publication. Future directions include testing this setup against other matrix sign function approximations and in dynamical fermion settings.

        \acknowledgments

            Computations were performed on the JUSUF and JURECA clusters at the Jülich Supercomputing Centre (JSC) under the VSR grant ``overlap reloaded''. S.G.\ and G.K.\ acknowledge support by the projects PulseQCD, HyperON, and Baryon8 (EXCELLENCE/0524/0269, VISION ERC-PATH 2/0524/0001, POSTDOC/0524/0001) co-financed by the ERDF and the Republic of Cyprus through the Research and Innovation Foundation. G.K.\ acknowledges partial support by AQTIVATE (Marie Skłodowska-Curie Grant Agreement No.\ 101072344).

\end{document}